\documentclass[rmp, twocolumn, nofootinbib]{revtex4}

\usepackage{epsfig} 

\newcommand{\beq}{\begin{equation}}
\newcommand{\eeq}{\end{equation}}
\newcommand{\beqa}{\begin{eqnarray}}
\newcommand{\eeqa}{\end{eqnarray}}
\newcommand{\om}{\Omega_m}
\newcommand{\omw}{\Omega_w}

\newcommand{\lam}{\Lambda}

\newcommand{\ls}{\mathrel{\raise0.27ex\hbox{$<$}\kern-0.70em \lower0.71ex\hbox{{
$\scriptstyle \sim$}}}}

\begin{document} 

\title{The Dynamics of Quintessence, The Quintessence of Dynamics} 
\author{Eric V. Linder}
\affiliation{Berkeley Lab, University of California, Berkeley, CA 94720, USA} 

\date{\today}

\begin{abstract} 
Quintessence theories for cosmic acceleration imbue dark energy with a 
non-trivial dynamics that offers hope in distinguishing the physical 
origin of this component.  We review quintessence models with an emphasis 
on the dynamics and discuss classifications of the different physical 
behaviors.  The pros and cons of various parameterizations are examined 
as well as the extension from scalar fields to other modifications of the 
Friedmann expansion equation.  New results on the ability of cosmological 
data to distinguish among and between thawing and freezing fields are 
presented. 
\end{abstract} 


\maketitle

\section{Introduction \label{sec:intro}}

Understanding the acceleration of the cosmic expansion is a landmark 
problem in physics, impacting gravitation, high energy and quantum 
physics, and astrophysics, 
and likely to revolutionize one or more of these fields.  The direction 
in which to look for a solution is almost wholly unknown currently.  Though 
there is no shortage of suggestions, most are far from a first principles 
explanation of how such physics arises. 

Perhaps the simplest proposal -- Einstein's cosmological constant $\Lambda$ 
\citep{einstein} -- is correct, though even so we 
have as yet no understanding of why it would arise, with the magnitude 
needed to explain acceleration occurring near the present epoch.  That 
puzzlement can be broken into two severe problems 
\citep{weinberg,carroll,padma}: the fine tuning problem 
of how $\lam$ appears with a magnitude (energy density or energy scale) 
so far from the natural (Planck) scale defined by fundamental constants, 
and the coincidence problem of why acceleration appears in our recent past, 
at a cosmic scale factor within 2 of the present value out of perhaps 
$10^{28}$ since inflation.  The cosmological constant is addressed in 
far greater detail in the articles by \citet{padmagrg,bousso} in this 
special volume. 

To paraphrase Winston Churchill speaking about democracy, it may be 
that the cosmological constant is the worst form of accelerating 
physics, except for all those other forms that have been tried from time 
to time.  Nevertheless, this article addresses those other forms, 
specifically dynamical physics that aims to ameliorate the coincidence, 
and/or fine tuning, problems.  We concentrate on the dynamics, 
the time evolution of the cosmological expansion physics, (mostly) from 
a canonical scalar field, given the name quintessence. 
See \citet{copeland} for a particle physics perspective. 

Section \ref{sec:history} provides a brief historical perspective on 
the development of quintessence theories.  Section \ref{sec:dynwwp} 
reviews key elements of the dynamics of quintessence and the physical 
origins of structure in the phase space, defining classes of models.  
Efficient representation of the dynamical behavior through parameterization 
or principal component analysis is discussed in Section \ref{sec:dynpar}, 
and we investigate in detail thawing models, those which approach 
cosmological constant behavior, in Section \ref{sec:thaw}.  In 
Section \ref{sec:extend}, we consider a selection of dynamical models 
beyond standard quintessence, and briefly mention the effects of expansion 
dynamics on growth of cosmic structure in Section \ref{sec:dyngro}.  
We conclude in 
Section \ref{sec:concl}.

\section{Origins of Quintessence \label{sec:history}} 

The role of a dynamical scalar field for recent acceleration of 
the cosmic expansion certainly owes a debt to the use of rolling 
scalar fields for early universe inflation. 
A scalar field, and more generally a negative equation of state, were 
implemented as a substitute for the cosmological constant in a flurry of 
activity in the 1980s.  On the theoretical side, 
\citet{linde86} proposed a simple extension from the flat potential of 
the cosmological constant to a tilted, linear potential, that releases 
the field to roll when the expansion rate of the universe decreases 
sufficiently, what is now called a thawing field.  In 1988, two nearly 
simultaneous papers by \citet{wett88} and 
\citet{ratrap} described in more detail cosmology in the presence of a 
quintessence field. 

At the same time, considerable work on the phenomenology of energy 
density components with an arbitrary (including negative) pressure 
to density, or equation of state, ratio was being carried out. 
\citet{wag86} discussed 
such generalized cosmology, and \citet{lin88s} then followed 
up on this with detailed investigation of a variety of cosmological probes 
of additional components with arbitrary equation of state.  These 
included tests of the expansion dynamics through distance, age, volume, 
and abundance measurements.  Particular attention was paid to light 
propagation in such a generalized cosmology, including possible 
inhomogeneities in the components \citep{lin88l} (some results 
occurred earlier in the unpublished thesis of \citet{kayser}).  
General equations of state had been considered in a formal way 
for the growth of structure within linear perturbation theory by 
\citet{kodamas}. 
Implications of general equations of state for growth were presented 
in \citet{fry} and \citet{linmpa}. 

Thus high energy physics theory and cosmology were all ready in the 
1980s for data exploring the expansion and growth histories of the 
universe.  It took another 10 years for observations \citep{riess98,perl99} 
to make the 
astonishing breakthrough that turned these speculations into a 
central subject of research into our understanding of gravitation, 
quantum physics, cosmology, and the fate of the universe.

\section{The Quintessence of Dynamics \label{sec:dynwwp}} 

\subsection{Scalar Field Basics} 

If we view the cosmological constant as a quantum zeropoint energy 
corresponding to the ground state of harmonic modes of a field filling 
space, we can picture this as an array of identical springs, motionless and 
each stretched to the same length.  By contrast, a scalar field would 
be a dynamical version of this, with the springs oscillating in time and 
having different lengths at different points in space.  That is, a scalar 
field is a very simple quantity, a magnitude at each point in space.  One 
can literally picture it as a field: a field of grass where each stalk 
may have been mown to a different height (a vector field could then be 
a field of trampled grass, where each stalk has a length and a direction 
in which it lies).  

For quintessence, we take a scalar field $\phi$ minimally coupled, i.e.\ 
feeling only gravity, passively through the spacetime curvature, and 
a self-interaction described by the scalar field potential $V(\phi)$.  
Moreover, we consider the kinetic contribution to the Lagrangian (the 
``bouncing of the springs'') to be canonical, i.e.\ involving only a 
term linear in the kinetic energy of the field.  (We briefly discuss 
relaxing these conditions in \S\ref{sec:extend}.)  
So the Lagrangian is about as simple as possible: 
\beq 
{\mathcal L}_\phi=\frac{1}{2}\partial_\mu\phi\partial^\mu\phi-V(\phi). 
\eeq 

Through the Noether prescription we define an energy-momentum tensor 
\beq 
T_{\mu\nu}=\frac{2}{\sqrt{-g}}\frac{\delta(\sqrt{-g}{\mathcal L})}{\delta 
g^{\mu\nu}}, 
\eeq 
where $g_{\mu\nu}$ is the metric and $g$ its determinant.  Comparing the 
result for a homogeneous and isotropic spacetime to the perfect fluid 
form allows identification of the energy density and pressure: 
\beqa 
\rho_\phi&=&\frac{1}{2}\dot\phi^2+V(\phi)+\frac{1}{2}(\nabla\phi)^2 \label{eq:rhophi} \\ 
p_\phi&=&\frac{1}{2}\dot\phi^2-V(\phi)-\frac{1}{6}(\nabla\phi)^2. \label{eq:pphi}
\eeqa 
Because late time acceleration requires a very light scalar field, with 
effective mass of order the Hubble parameter, the Compton wavelength of 
the field will be of order or larger than the Hubble scale and so the 
field is expected to be spatially smooth within the Hubble scale.  Therefore 
we neglect the spatial gradient terms in the energy density and pressure. 
These quantities can be put into the usual Friedmann equations to solve 
for the expansion history of the scale factor vs.\ time, $a(t)$, from 
the Hubble parameter $H=\dot a/a$ and acceleration $\ddot a$. 

Because both the energy density and pressure enter the equations, it 
is convenient to define an equation of state ratio, 
\beq 
w=p_\phi/\rho_\phi, 
\eeq 
which is generally time varying.  When we refer to dynamical fields, 
we generally mean time-varying $w$, i.e.\ $w\ne {\rm constant}$.  
(Although the 
energy density of constant $w$ models varies with time, this happens as 
well with matter or a frozen network of cosmic strings, say, and so does 
not capture the flavor of ``dynamics''.) 

The equation of motion for the scalar field is the Klein-Gordon equation 
\beq 
\ddot\phi+3H\dot\phi=-dV/d\phi, \label{eq:kg}
\eeq 
and is interchangeable with the continuity equation.  For example, 
multiplying through by $\dot\phi$ gives the sequence 
\beqa 
[\dot\phi^2/2]\,\dot{}+6H[\dot\phi^2/2]&=&-\dot V \\ 
\dot\rho_\phi-\dot V+3H(\rho_\phi+p_\phi)&=&-\dot V \\ 
\frac{d\rho_\phi}{d\ln a}=-3(\rho_\phi+p_\phi)&=&-3\rho_\phi\,(1+w). \label{eq:dotrho} 
\eeqa 
where we have turned Eqs.~(\ref{eq:rhophi})-(\ref{eq:pphi}) around to use 
\beqa 
V&=&(\rho_\phi-p_\phi)/2=\rho_\phi(1-w)/2 \label{eq:vw} \\ 
K&\equiv&\dot\phi^2/2=(\rho_\phi+p_\phi)/2=\rho_\phi(1+w)/2. \label{eq:kinw} 
\eeqa 

From the above equations we can formally go back and forth from the 
field description to the fluid description or equation of state.  
From Eqs.~(\ref{eq:rhophi})-(\ref{eq:pphi}) we see that 
\beq 
w=\frac{K-V}{K+V}, \label{eq:wkv}
\eeq 
so for some specified theory we can calculate the equation of state 
and then the effects on the cosmological expansion.  The other direction, 
starting from observations of the cosmological expansion, is slightly 
more complicated: 
\beqa 
\rho_\phi(a)&=&\Omega_w\,\rho_c\,\exp\left\{3\int_a^1 d\ln a\,[1+w(a)]\right\} \label{eq:rhoa} \\ 
\phi(a)&=&\int d\ln a\,H^{-1}\sqrt{\rho_\phi(a)\,[1+w(a)]} \\ 
V(a)&=&\rho_\phi(a)\,[1-w(a)]/2 \\ 
K(a)&=&\dot\phi^2/2=\rho_\phi(a)\,[1+w(a)]/2. 
\eeqa 
Such reconstruction of the scalar field physics is made difficult by 
a number of issues: noisiness of measurements of the expansion, translation 
from the measured quantity to density or equation of state through one 
or two derivatives, and finite range of scale factor, or redshift 
$z=a^{-1}-1$, coverage.  In particular, from the last of the equations 
above we see that 
\beq 
\dot\phi=[\rho_\phi(1+w)]^{1/2}\lesssim HM_P\,(1+w)^{1/2}, 
\eeq 
so for cases when $1+w\ll 1$ (as seems to be implied by observations), 
only a small region of the scalar field physics, $\Delta\phi\sim 
\dot\phi/H\ll M_P$, can be probed.  All these issues together makes 
reconstruction problematic, and we do not consider it further.  (For 
attempts to carry it through, see \citet{sahnistaro} and references 
therein.) 

While we cannot reconstruct in detail the scalar field potential, we 
can derive considerable insight into the accelerating physics from study 
of its dynamics.  We can guess from the spring picture at the beginning 
of this section that there will be at least three basic quantities we 
want to know: how much energy is there in the field, how springy is it, 
and how stretchy are the springs?  The energy density $\rho_\phi$ is 
conveniently written in terms of the dimensionless density 
$\omw=\rho_\phi/\rho_c$, where $\rho_c=3H_0^2/(8\pi G)$ is the critical 
density.  For a spatially flat universe, $\omw=1-\om$, where $\om$ is 
the dimensionless matter density.  The analog of the springiness is 
how spacetime curvature reacts to the accelerating component; the 
passive gravitational mass is given by $\rho+3p$, with acceleration 
induced by a component possessing $p<-\rho/3$, or $w<-1/3$.  So we can 
regard $w$ as a measure of the springiness.  As the universe expands, 
the springs react, changing their springiness, like stretching the coils 
of a spring.  This time variation can be taken as $w'=dw/d\ln a=\dot w/H$. 
Thus we are primarily interested in $\omw$, $w$, $w'$.  The last two 
quantities give a phase space for the dynamics which we will see is 
enlightening.

\subsection{General Dynamical Behavior} 

Scalar fields can at any epoch have one of four behaviors. 
Their rolling can be fast, slow, more or less steady, or oscillatory. 

{\it Fast roll:\/} Fast rollers have kinetic energy exceeding their 
potential energy, and so by Eq.~(\ref{eq:wkv}) have $w>0$.  These 
clearly do not act to accelerate 
the cosmic expansion, but a fast roll epoch (``kination'') is a 
characteristic of tracker models, which follow attractor trajectories in 
their dynamics such that at certain epochs their equation of state is 
determined by the dominant 
energy density component of the universe.  Because of the fast roll, 
the scalar field can rapidly decrease its energy density from an initial, 
early universe value near the ``natural'' Planck scale to a much smaller 
value that will make it suitable for the observed present energy density. 
Due to the attractor solution for the dynamics, for certain forms of 
the potential, there is a large variety of initial conditions -- ``basin 
of attraction'' -- that can deliver a reasonable present energy density, 
thus addressing the fine tuning problem of the cosmological constant. 
Of course the field must leave both the fast roll regime and the tracking 
regime if it is to cause acceleration and dominate the energy density, 
so the coincidence problem is not 
completely solved.  In particular, tracking fields have difficulty reaching 
equations of state $w\lesssim-0.7$, in tension with observations, and so 
are no longer considered front runners for explaining the acceleration. 
For more on trackers (and the earlier ``tracers''), see 
\citet{zlatev99,swz99,ferreirajoyce,liddlescherrer}.  

{\it Slow roll:\/} When the kinetic energy is much smaller than the 
potential energy, the equation of state is strongly negative, $w\approx-1$. 
Of course this only leads to acceleration of the expansion if the dark 
energy also dominates the energy density.  The field is nearly frozen, 
and the dark energy density is nearly constant (while matter and radiation 
are rapidly diluting due to the expansion), so it would eventually 
come to dominate the universe if nothing else changed.  Note that because 
matter is not negligible, even today, a field we think of as slowly 
rolling, $w\approx-1$, may well not have a small value for $V'/V$ 
(see, e.g.~\citet{paths}), which is a conventional slow roll parameter 
for inflation 
(where the accelerating component is completely dominant).  Quintessence 
models that always have the potential dominating over the kinetic term 
encounter the same fine tuning and coincidence problems as the cosmological 
constant, lacking the basin of attraction of tracker models.  Thus 
generically, we want a combination of fast and slow roll behavior for a 
successful model. 

{\it Steady roll:} Referring to the original quintessence 
model of \citet{linde86} using a linear potential, this category is somewhat 
of a misnomer since the field does have fast and slow roll epochs over 
its entire history.  However, the linear potential model does have a 
constant right hand side of the Klein-Gordon equation of motion, and 
for a long time the dynamics stays reasonably close to the line where 
the field acceleration $\ddot\phi$ (not the cosmic acceleration $\ddot a$) 
is zero (see \S\ref{sec:phase} below).  This model is not only the simplest 
generalization of the cosmological constant but is also interesting in 
its overall history.  It starts generically from a frozen, cosmological 
constant-like state due 
to Hubble friction, then thaws and rolls down the potential.  However, 
because the potential has no minimum, the field rolls into territory 
where the potential goes negative, which actually leads to a collapsing 
universe, rather than an accelerating expansion.  These models therefore 
have a finite future history, with a ``doomsday time'' 
\citep{kratoch1,kratoch2}. 

{\it Oscillation:}  Common potentials in renormalizable field theories 
include $V(\phi)\sim\phi^n$, which have a minimum for $n$ even.  While 
the field will have a conventional rolling stage, eventually it will 
reach the minimum and oscillate around it.  If the period for oscillation 
is much smaller than the Hubble time (as is generally the case) then 
the effective equation of state becomes \citep{turner83} 
\beq 
w=\frac{n-2}{n+2}. 
\eeq 
For a quadratic potential, the field acts like nonrelativistic matter, 
and for a quartic potential it acts like radiation.  

One intriguing 
example of such a field is the axion, or more generally pseudo-Nambu 
Goldstone bosons (PNGB).  If we consider them during the regime when 
they are still rolling rather than oscillating, they can accelerate 
the expansion, though this acceleration will eventually fade away as 
the field evolves to its oscillatory, matter-like phase \citep{frieman95}.  
PNGB potentials are also radiatively stable against quantum corrections, 
unlike an ad hoc $V(\phi)$ that might be written down but then acquire 
a non-zero ground state (cosmological constant) and distortion of its 
shape.  Thus the physics of such pseudoscalar fields offers some promise 
for a fundamental, high energy physics origin rather than merely a low 
energy effective potential.  The PNGB potential looks like 
\beq 
V(\phi)=V_0\,[1+\cos(\phi/f)], 
\eeq 
where $f$ is a symmetry energy scale.  Because the potential is 
nonmonotonic and the slope of the potential changes from concave to 
convex, a number of interesting effects can arise, such as 
mimicking super-negative equations of state $w<-1$ and nontrivial dynamics 
\citep{ljhall,kalopersorbo,csaki}. 
For a complex field, one has degrees of freedom in both the modulus 
and the phase, and researchers have considered making one act as dark 
energy and the other as dark matter \citep[e.g.][]{pngbmat}, or one giving 
recent acceleration and one early universe inflation 
\citep[e.g.][]{rosenfeldfrie}.  
Other elaborations include spintessence \citep{guhwang,spintess}.

\subsection{Fundamental Modes of Dynamics \label{sec:phase}} 

While in the previous subsection we considered the behavior 
of the scalar field dynamics at any one moment, 
considerably more insight comes from investigating the 
overall dynamical history given by the trajectory through 
phase space. In particular, we will be interested not only 
in the present characteristics, but the asymptotic past 
and future states. 

By examining the physical impact of the three different terms 
in the Klein-Gordon equation (\ref{eq:kg}) we can identify boundaries 
in the phase space corresponding to different physical 
conditions. 

\begin{itemize} 

\item {\bf Phantom line:} This line separates physics obeying the 
null energy condition \cite{hawkingellis}, $\rho+p\ge0$ ($w\ge-1$), 
from physics violating it.  Also, consider the friction term $3H\dot\phi$. 
From Eq.~(\ref{eq:kinw}) one sees that where the sign of this 
term changes, i.e.\ $\dot\phi=0$ as the field stops rolling 
in one direction (and possibly begins rolling in another), 
corresponds to 
\beq 
w=-1.
\eeq 
Canonically the field has $w\ge-1$ but there are various mechanisms 
(see \S\ref{sec:extend}) for achieving $w<-1$, what is referred to 
as the phantom regime \citep{phantom}. 

\item {\bf Null line:} Consider the forcing term of the potential 
slope.  When the field rolls down the potential, $\dot V\le0$, 
this corresponds to 
\beq 
w'\ge -3(1-w^2), 
\eeq 
where we have used Eqs.~(\ref{eq:dotrho})-(\ref{eq:kinw}) 
to convert the variables $\dot V$ and $\dot\phi$ to $w$, $w'$. 
If the field 
has a (noncanical) negative kinetic energy so it rolls up the potential 
then the inequality flips but at the 
same time the sign of $w$ changes so $w<-1$ (one can 
think of this as the energy density increasing with 
time, following Eq.~\ref{eq:dotrho}). 
Thus the null line passes 
smoothly through the point $(w,w')=(-1,0)$. 

\item {\bf Coasting line:} Consider the acceleration term $\ddot\phi$. 
Generally, at late times, the field accelerates due to the potential 
forcing dominating over the friction, or decelerates 
if the friction dominates over the potential slope 
(note this should not be confused with the acceleration 
of the cosmic expansion, which holds in either 
case if $w$ is sufficiently negative). Again from 
Eq.~(\ref{eq:kinw}) the dividing line between these dynamics, 
where the field is freely coasting at constant velocity 
$\dot\phi$, is 
\beq 
w'=3(1+w)^2, 
\eeq 
with $w'$ greater (smaller) than this for field acceleration 
(deceleration). 

\end{itemize}

These three boundaries give general physical divisions 
for the dynamical behavior of the field. The general equation 
relating the phase space variables can be derived by 
taking the derivative of Eq.~(\ref{eq:vw}) 
and using the continuity 
equation (\ref{eq:dotrho}) to obtain 
\beq 
w'=-3(1-w^2)-(1-w)(1+w)^{1/2}\sqrt{\frac{3\Omega_w(a)}{8\pi}} 
\frac{M_P V_{,\phi}}{V}. 
\eeq 
We can readily verify the null line corresponds to $V_{,\phi}=0$ 
(and one can specialize to the coasting line with a little 
more effort). These conditions were defined in \citet{caldlin} 
and developed further in \citet{scherrerwwp} and \citet{paths}. The last 
reference in particular goes into more detail about the derivation 
and the effect of the ratios of different terms in the Klein-Gordon 
equation, as well as ``slow roll'' parameters of the potential. 

Without the need for quantitative analysis of the ratios 
of Klein-Gordon terms, one can broadly understand 
the dynamics by examining the relative dominance of the 
driving vs.~dragging terms, following \citet{caldlin}. If the Hubble
friction dominates at early times, then the field will 
be pinned and act like a static cosmological constant. 
As the cosmic expansion reduces the Hubble parameter, 
eventually the potential slope induces the field to begin 
rolling: such models are said to be {\it thawing\/}, and their 
dynamics in phase space shows them ``leaving $\lam$'', moving to 
less negative $w$ with positive $w'$.  In particular, fields 
that thaw during the matter dominated epoch leave $\lam$ along 
the track $w'=3(1+w)$.\footnote{Fields whose initial conditions 
$\dot\phi_i$ are fine tuned can avoid this.  Also, if the potential driving 
term is very large, for example in PNGB fields with symmetry 
energy scale $f\ll M_P$, then one can have $w'>3(1+w)$.} 
As the matter domination wanes, the trajectory will curve according 
to the driving force from the potential slope; since the potential 
(eventually) becomes less steep as it approaches the 
minimum, the field acceleration decreases and the curve 
is toward the coasting line, i.e.~smaller $w'$. For broad 
classes of potentials the condition that dark energy not 
completely dominate the energy density of the universe 
by the present means that thawing fields are still accelerating 
along the potential and the dynamics has a lower 
bound roughly given by $w'>1+w$ (for $\Omega_w<0.8$ and 
$w<-0.8$). Thus the thawing region of phase space is 
defined by a dynamical history 
\beq 
1+w\lesssim w' \le 3(1+w). \label{eq:thawlim}
\eeq 

The alternative is that the potential forcing dominates 
over the Hubble drag at early times, i.e.~the potential 
is sufficiently steep to overcome the friction from cosmic 
expansion. Such fields will look different from the cosmological 
constant at early times. Certain forms of potential 
possess special attractor properties, as discussed in the 
previous subsection, that during the matter dominated 
epoch cause the scalar field dynamics to have a constant 
equation of state determined by the background expansion. 
As the dark energy density becomes relatively more 
important, these fields will depart from their tracking behavior 
and roll according to the dynamics of their potential. 
As the field rolls toward the minimum, decelerating 
in its motion (lying below the coasting line), gradually 
approaching asymptotically a static cosmological constant 
state, it is said to be {\it freezing\/}. In its ``approaching 
$\lam$'', the field contributes an energy density $\rho_w\sim H^{2(1+w)}$, 
but \citet{paths} showed that any $H^\alpha$ model approaching $w=-1$ 
does so along the asymptotic trajectory $w'=3w(1+w)$. 
Conversely, since dark energy dominates (though not 
fully) today, the field must have departed its matter dominated 
tracking behavior and moved some distance 
away from the constant $w$ line. For broad classes of potential 
this leads to a present value $w'\lesssim 0.2w(1+w)$ (for 
$\omw>0.6$ and $w<-0.8$). Thus the freezing region of 
phase space is defined by a dynamical history 
\beq 
0\le w'\le 3w(1+w), \label{eq:freezlim} 
\eeq 
with the present value of $w'$ more tightly restricted. 

\begin{figure}[!htb]
\begin{center} 
\psfig{file=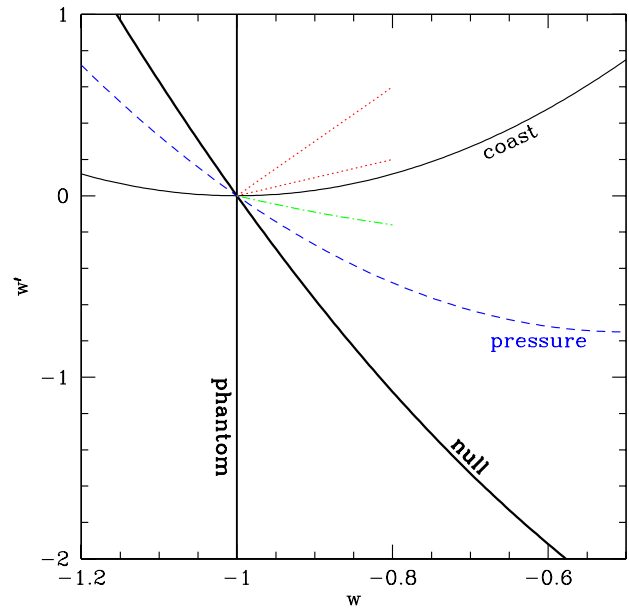,width=3.4in} 
\caption{The dynamical phase space $w$-$w'$ is divided by three curves 
defined by physical conditions: the phantom line $w=-1$, the null line 
$w'=-3(1-w^2)$ following from a flat potential, and the coasting 
line $w'=3(1+w)^2$ following from constant field velocity.  These extend 
across the phase space.  In addition, canonical dynamics leads to the distinct 
regions of the thawing regime bounded by the red dotted lines and the 
freezing regime bounded between the green dot-dashed curve and the blue 
dashed curve (the latter given by the constant pressure condition). 
}
\label{fig:phase} 
\end{center} 
\end{figure}

Figure \ref{fig:phase} illustrates the three critical dividing 
lines of the phantom, null, and coasting curves in the dynamical 
phase space.  In addition it shows the upper and lower boundaries 
of the thawing and freezing regions.  Note that the lower boundary 
of the freezing region coincides with the constant pressure curve 
(with an adiabatic sound speed $c_a^2=0$) discussed in \S\ref{sec:extend}.

Comparing Eqs.~(\ref{eq:thawlim}) and (\ref{eq:freezlim}), we see that 
they define 
narrow, distinct regions in the phase space where scalar 
field theories obeying a combination of theoretical and 
observational conditions lie. In particular, there are fairly 
strongly physically motivated outer boundaries defining 
the extremes of $w'$. The exact inner boundaries are more 
a function of empirical constraints on the present expansion, 
but there is a distinct intermediary zone unfavorable 
for habitation. This ``desert'' lies around the 
coasting line: only highly fine tuned models would, after 
the many e-folds of cosmic expansion influencing the 
scalar field equation of motion, find themselves almost 
perfectly balanced between field acceleration and deceleration, 
$\ddot\phi\approx0$. 

Two important implications of the physical division 
into distinct thawing and freezing regions are for the questions 
of observationally distinguishing dynamical dark 
energy from $\lam$ and distinguishing the physical origin of 
the dark energy (e.g.~field theories with thawing vs.~freezing 
characteristics). Because of the degeneracy directions 
of essentially all cosmological probes (see the articles by 
\citet{leibundgut} and \citet{nichols} 
in this volume), the entire thawing region is 
difficult to distinguish from the cosmological constant if 
the data is only at the sensitivity level of a constant, or 
time averaged, $w$. For example, the entire thawing region 
would give an apparent $\langle w\rangle\approx-1\pm0.05$. Thus 
experiments sensitive to $w'$ are necessary for deciding between 
this half of the dynamical phase space and the cosmological 
constant. For distinguishing between the classes of 
effective field theories, one would like to have cosmological sensitivity 
to the time variation of $\sigma(w')\lesssim 2(1+w)$ to resolve the 
separation between the thawing and freezing regions. For 
indepth discussion of mapping the cosmic expansion history, see 
the review article by \citet{linrop}.

\subsection{More Complicated Dynamics \label{sec:complex}} 

In the previous subsection we gave physical motivations 
for bounded regions in phase space but we emphasize 
these are based on a combination of generic behavior 
and empirical data, not an absolute exclusion of other 
possible behaviors. In particular, they relied on a standard 
matter dominated epoch at high redshift, canonical 
scalar fields, avoidance of fine tuned initial conditions and 
potential shapes, and ``fundamental modes'' of dynamics. 
We discuss extension of the dynamics to beyond canonical 
scalar fields in \S\ref{sec:extend}; here we consider initial conditions 
and fundamental modes. 

Initial conditions on the scalar field dynamics are quite  
important, e.g.~one could consider a field so perfectly 
balanced on a maximum of its potential that it only starts  
rolling yesterday, or a field that has recently passed a 
minimum of its potential and is now rolling uphill, or a 
field with kinetic and potential energies exactly crafted 
so the dynamics is missing (constant equation of state) or 
is coasting. Physics does not forbid any of these a priori, 
but our sense of naturalness disfavors them. If dynamical 
conditions are set by hand at recent times, rather than 
the field settling into an evolution following its equation 
of motion over many e-folds in the early universe and 
then a matter dominated epoch, then virtually arbitrary 
behavior can result \citep{hutpeir,liholz}. One could fine tune the field 
such that one does not extract general physical precepts 
on the dynamics, but rather the phase space trajectories 
would spell out your name. 

Under the physics of field evolution through the cosmic 
expansion history, including a matter dominated epoch, 
the phase space structure described in the previous subsection 
generically holds. One further necessary ingredient 
is that we are talking about fundamental modes, or 
``atoms'', of the dark energy -- the quintessence of dynamics. 
If one combines multiple elements together, such as a 
scalar field plus a cosmological constant, or plus matter, 
or plus another scalar field, then one can indeed break the 
physical boundaries (just as multifield inflation can break consistency 
relations and other basic predictions). That is, the phase space structure 
applies to the dynamics of a single, fundamental field, 
not an effective field of multiple origins. 

We can investigate this further by examining the effect on the 
equation of state when multiple elements are combined.  For the 
simplest approach, we consider adding together two components: 
a canonical scalar field plus either a cosmological constant, 
a matter component (e.g.~misestimation of $\om$ or dark energy 
contribution to dark matter), or another scalar field. 

The effect of combining two such noninteracting components is 
given by an effective dynamical equation of state 
\beq 
w_{\rm eff}=w_1\frac{\delta H_1^2}{\delta H_1^2+\delta H_2^2}+ 
w_2\frac{\delta H_2^2}{\delta H_1^2+\delta H_2^2}, 
\eeq 
where $\delta H_i^2$ is the contribution of component $i$ to the 
Friedmann equation.  This approach was used to first point out 
phantom crossing, evolution across $w=-1$, by two scalar fields 
\citep{lingrav} (also see \citet{hucross}).  The dynamics is affected as 
\beqa 
w_{\rm eff}'=3w_{\rm eff}(1&+&w_{\rm eff})+\frac{\delta H_1^2}{\Sigma} 
[w_1'-3w_1(1+w_1)] \nonumber\\ 
&+&\frac{\delta H_2^2}{\Sigma} [w_2'-3w_2(1+w_2)], 
\eeqa 
where $\Sigma=\delta H_1^2+\delta H_2^2$.  Note that two constant 
pressure components (where $w_i'=3w_i(1+w_i)$) add without affecting 
the dynamics.  In particular, any combination of matter plus $\lam$ 
keeps the same trajectory, just moving the position along the track. 

Furthermore, this formula implies that the sum of components, each of 
which lies on the same side of the curve $w'=3w(1+w)$, has effective dynamics 
doing likewise.  For example, two kinetic k-essence components give an 
effective dynamics that is still kinetic k-essence-like.  Similarly, the null 
condition $w'>-3(1-w^2)$ cannot be overcome by summing components obeying 
$w_i'>-3(1-w_i^2)$.  Other than respecting these two boundaries, the 
dynamics can change significantly on combining components. 

\begin{figure}[!htb]
\begin{center} 
\psfig{file=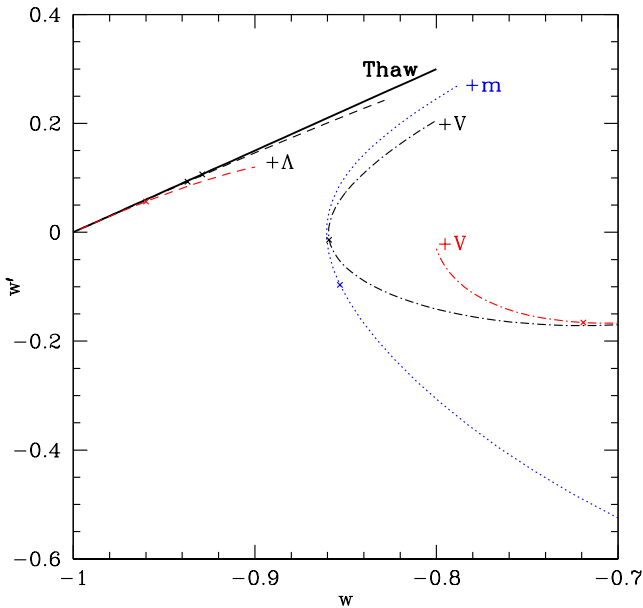,width=3.4in} 
\caption{Dynamics involving combination of physics can violate the 
fundamental phase space regions.  To the original thawing scalar 
field trajectory (solid black), we add a cosmological constant ($+\lam$), 
extraneous matter or quartessence component ($+m$), or freezing 
scalar field ($+V$).  We fix $w_0=-0.8$ for the 
fields and take the total dimensionless dark energy density to be 0.7.  
For the second component of $\lam$ or $V$ we take 
$\Omega_2=0.1$ (darker, black) or 0.35 (lighter, red); for included 
matter $\Omega_{+m}=0.01$.  Curve endpoints correspond to $z=0$, with 
x's at $z=1$. 
}
\label{fig:nonfund} 
\end{center} 
\end{figure}

To an initial thawing scalar field we add either a cosmological 
constant component, a matter component, or a freezing field.  
Figure~\ref{fig:nonfund} shows that such combinations, as opposed to 
the fundamental modes or ``atoms'' we discussed in the previous 
subsection, do not adhere to the restricted thawing and freezing 
regions of the phase space.  Convolutions of different physics can 
drastically differ from those fundamental behaviors.

Adding a freezing field to a thawing field dramatically alters the 
trajectory, since at early times the freezing field will dominate. 
(Adding extra components to a dominant freezer has less effect.) 
The phase space tracks therefore start off in the freezing regime 
but curve up toward the thawing regime, possibly lying today in the 
desert region between the two regimes.  
A cosmological constant rotates 
the dynamics toward $w'=0$ and draws it in toward $w=-1$ (see also 
\citet{caldlin}); this does 
not generally move a thawing field out of the thawing region.  Including 
a matter like component with the thawing field has the most severe 
effect.  Adding a mere $0.01$ in dimensionless matter-like energy density 
alters the track wildly -- this points up strongly the dangers in 
attempted direct reconstruction of the dynamics from $H(z)$ or the 
distance-redshift relation.  Misestimation of 
$\om$ by $0.01$ will completely distort the true dark energy dynamics.

\section{Describing the Dynamics \label{sec:dynpar}}

The phase space dynamics discussed in the previous 
section presents the dark energy physics in terms of 
a function $w(a)$ and its derivative $w'$, describing the 
``springiness'' and ``stretchiness'' of the spacetime in reaction 
to the dark energy. Each theoretical model presents 
its particular description of the function and we can check 
each against the data to determine whether the model 
fits. However, there are $10^x$ theoretical forms (potentials 
or equation of state functions) already postulated, each with their 
own parameters. Moreover, we would like to predict the 
results of experiments, or design experiments, more generally 
than for a given theory or set of existing theories. 

This shows the need for a model independent approach, 
based on a parametrization of the equation of 
state function or a similar quantity. Because we want the 
parametrization to stay close to the underlying physics, 
of which both the dark energy density and pressure enter, 
we concentrate on the pressure to density ratio, or 
equation of state ratio. However parametrization of other 
quantities such as distances, Hubble parameter, or density 
alone have been considered (see, e.g., \citet{sahnistaro} and 
references therein); two cautions should be 
stated about this route: certain forms bias the extraction 
of the underlying physics, see e.g.~\citet{jonsson,linbias}, and 
if one eventually 
wants the equation of state then one is forced to 
take numerical derivatives of a quantity extracted from 
noisy data. 

Numerous parametrizations exist for the equation of 
state $w(a)$ but the vast majority are purely ad hoc. We 
here consider a very few that are phenomenological in the 
best sense, i.e.~generalized from the behavior of physically 
motivated sets of models. 
From the previous section we have 
seen that a single parameter model, i.e.~$w={\rm constant}$, involves 
highly fine tuned physics to remove the dynamics. 
While one way out of this is to invoke a physical symmetry, 
such as a topological defect origin, which can produce 
$w=-N/3$ for a frozen network of $N$-dimensional 
defects (e.g.~$N=2$ domain walls \citep{domain} or $N=1$ light 
cosmic strings \citep{strings}), such values are not consistent with 
data. 

This leads us to two parameter models as the next 
simplest alternative. The parametrization 
\beq 
w(a)=w_0+w_a(1-a), 
\eeq 
where $w_0$ is the value today ($a=1$) and $w_a$ is a measure 
of the time variation $w'$, is 
widely used in the literature. 
It is important to realize 
that it is in no way a mathematical expansion about 
the present: neither its important introduction by \citet{chevallier} 
nor the physical foundation work by \citet{linprl} employed 
a Taylor expansion, nor would that be mathematically 
convergent. Therefore $w_a$ is not an expansion parameter 
about $z=0$, but rather a fit parameter describing the 
overall time variation $w'$. The original convention \citep{linprl} 
giving the best description is 
\beq 
w_a\equiv (-w'/a)|_{z=1}=-2w'(z=1). 
\eeq 

\citet{linprl,linidm} give several physical supports for the $w_a$ 
parametrization: 1) excellent approximation 
to the exact field equations for a broad range of 
fundamental or straightforward scalar field potentials, 
2) well behaved at both low and high redshift, 3) robust 
against bias, e.g.~if one extends the form to further 
parameters, the $w_0$, $w_a$ parameter values estimated 
are not strongly affected, 4) model independence. For 
example, a SUGRA inspired model that evolves from 
$w(a\ll1)\approx-0.2$ to $w_0=-0.82$ -- a substantial variation 
-- has its equation of state reproduced to within 3\% back  
to $z=1.7$ and the distance-redshift relation in such a 
cosmology is accurately matched to 0.2\% back to CMB 
last scattering by $w_0=-0.82$, $w_a=0.58$. 

Of course a two parameter description cannot describe all 
possible dynamics; in particular it begins to break 
down for rapid transitions in the equation of state or 
oscillations.  
However, for the fundamental modes highlighted in the previous 
section it serves as an excellent, broad 
(i.e.~model independent, good for both thawing and freezing) 
parametrization of the physically favored dynamics. 

Another two parameter form, which is motivated from 
the energy density rather than the equation of state, is 
the bending parametrization of \citet{bending}. This was designed to 
describe early dark energy models where at high redshift 
(near the CMB last scattering surface, $z\approx10^3$) the 
scalar field component has nonnegligible energy density 
(though it is then acting in a decelerating, rather than 
accelerating, manner on the expansion, so it is not exactly 
dark energy). The bending form has 
\beqa 
\ln\frac{\omw(a)}{\om(a)}&\equiv&R_0-\frac{3w_0\ln a}{1-b\ln a} \\ 
w(a)&=&\frac{w_0}{(1-b\ln a)^2}, \label{eq:bend} 
\eeqa 
where $R_0=\ln(\om^{-1}-1)$ and $b$ is related to the early dark energy 
density. The dynamics of this parametrization 
is that in the past it approaches $w=0$, $w'=0$ (i.e.\ a finite dark 
energy density that acts like matter), at some future time $a_*=e^{1/b}$ 
it runs to $w=-\infty$, $w'=-\infty$, and then returns along the same 
trajectory to $w=0$, $w'=0$ in the further future.  The phase space track 
is defined by $w'=2bw_0\,(w/w_0)^{3/2}$.  At any given time 
in the past the variation must be slower than $w'=-(8/27)w_0/\ln a$.  

A generalization of the $w_a$ form to three parameters 
was put forward by \citet{wext}. This eases the property of the $w_a$ 
form where the parameter $w_a$ plays two roles: it describes 
the characteristic time variation $w'$ but it also determines 
the asymptotic past value of $w(a\ll1)\to w_0+w_a$. The 
extended form has 
\beq 
w(a)=\frac{w_p z+w_0 z_t}{z+z_t}, 
\eeq 
where $w_p$ is the asymptotic past value and $z_t$ is the transition 
redshift.  When $z_t=1$, this reduces to the $w_a$ parametrization. 
The phase space dynamics is a parabola from $(w,w')=(w_p,0)$ to 
$(w_f,0)$, crossing $w=-1$ if $w_0<w_p$. 

To describe a monotonic $w(a)$ which transitions 
smoothly from some asymptotic past value $w_p$ to some 
asymptotic future value $w_f$ requires a minimum of four 
parameters: $w_p$, $w_f$, the epoch of transition $a_t$, and a 
rapidity parameter $\tau$. (Note that the previous 
models are not bounded in the future; this is not 
overly worrisome because we have no data on the expansion 
future.) Such forms are particularly successful in 
describing tracking models which have both asymptotic 
past and future equations of state. The transition can be 
described by many functional forms, but the two most 
common four parameter equations of state both adopt 
``Fermi-Dirac'' transitions. The kink model \citep{corakink} 
takes this in scale factor $a$, obtaining 
\beq 
w(a)=w_0+(w_m-w_0)\frac{1+e^{a_t/\Delta}}{1-e^{1/\Delta}}\frac{1-e^{(1-a)
/\Delta}}{1+e^{(a_t-a)/\Delta}},
\eeq 
where $w_m$ is the asymptotic value in the matter dominated era and 
$\Delta$ is related to the rapidity, 
while the e-fold model \citep{lh05} does the transition in the 
expansion e-fold factor $\ln a$, obtaining 
\beq 
w(a)=w_f+\frac{w_p-w_f}{1+(a/a_t)^{1/\tau}}\,. 
\eeq 
One of the advantages of the e-fold model is that it allows 
an analytic expression for the Hubble parameter $H(a)$.  

One could continue developing more complicated forms 
but sadly even the next generation of experiments will 
not be able to constrain stringently more than two equation 
of state parameters \citep{lh05}.  This conclusion holds whether 
dealing with parameters per se or principal components (see below). 
Happily, the $w_a$ parametrization is quite satisfactory in giving 
a model independent, good approximation to the dynamics. 

Nevertheless, let us briefly consider principal component 
analysis (PCA). This approach attempts to 
gain some independence from the particular form of 
parametrization, letting the data define the best constrained 
combination of information. This is a valuable 
tool; see \citet{hutstark} for its development for the dark energy equation 
of state, and \citet{coohut} for an adaptation localizing the 
principal components in redshift. PCA has the advantage 
over parametric forms in its nonparametric flavor, and in specifying 
what a particular survey measures best, 
however its results are dependent on ingredients other 
than the underlying physics: the type of cosmological 
probe, the details of the data, the fiducial cosmology, 
and priors. That is, a principal component derived 
from one specific experiment is not exactly comparable 
to a principal component from another experiment, or 
the same experiment over a different redshift range. By 
contrast, $w_0$ and $w_a$, say, mean the same thing regardless 
of probe, survey, cosmology, or priors. (We are talking 
about the meaning of the variables, not the estimation of 
the fit values.) Thus, PCA is likely to be of most use as 
a complementary tool alongside parametric fits. 

Note there has been some confusion in the literature 
regarding the accuracy of PCA fits, with some claims that 
more than two principal components can be stringently 
fit by next generation experiments. In the analyses where 
there appear to be more than two well fit parameters, 
this arises from consideration only of low noise in the 
component coefficients $\alpha_i$, e.g.~$\sigma(\alpha_i)$, 
not high signal to noise criteria $\sigma(\alpha_i)/\alpha_i$. 

So we appear restricted to two parameters for our equation 
of state description. However, a tilt from the cosmological 
constant value, $1+w$, and a time variation, 
$w'$, contain rich information on the physics responsible 
for the acceleration of the universe. Given we have only 
two parameters, are we sure that $w_0$ and $w_a$ represent 
the best, model independent parameters? No, we have 
no guarantee of this and we should continually be on the 
lookout for improvements, though to date $w_0$, $w_a$ have 
served extremely well. 

One idea for an alternate parameter involves the so-called 
pivot or minimum variance equation of state $w_p$. 
This is the equation of state at the scale factor $a_p$ where 
the variance $\sigma^2(w(a))$ is minimized, i.e.~$w_p=w(a_p)$. 
Note that $w_p$ is also decorrelated with $w_a$, with zero covariance 
between their estimations, but this holds only 
due to the specific linear dependence of the equation of 
state $w(a)$ on $w_a$; generally the minimum variance value is not 
decorrelated with other equation of state parameters. The pivot 
parameter possesses many of the same issues as the PCA approach: 
lack of an invariant physical meaning due to dependence 
on probe, survey, model, and priors. It is sometimes 
useful however for the narrow question of whether 
the data are consistent with a cosmological constant cosmology 
(in one direction, at least; one can find $w_p=-1$ 
yet have dynamical dark energy).  Note for thawing 
models the deviation $1+w$ is greatest at $z=0$ so a 
parameter at $z_p$ may not be optimal even for this question. 
\citet{linbias} showed that generally $w_p$ is more subject to 
bias than either $w_0$ or $w_a$. 

Another suggestion for alternate parametrization involves 
either so-called statefinder variables ($r,s$) \citep{statefinder} 
or combinations of derivatives of the cosmic scale factor such 
as the deceleration parameter $q=-a\ddot a/\dot a^2$ and jerk 
$j=a^2\dddot a/\dot a^3$ \citep{jerk}.  Note that either parametrization 
convolves the equation of state parameters with the energy density: 
\beqa 
q&=&\frac{1}{2}+\frac{3}{2}w\omw(a) \\ 
j&=&1-\frac{3}{2}\omw(a)\,[w'-3w(1+w)]=q+2q^2-q'.
\eeqa 
($r$ is the same as $j$, and $s=[3w(1+w)-w']/(3w)=c_a^2(1+w)/w$, where 
$c_a^2$ is the adiabatic sound speed.)  These approaches also conflate 
different physics: $j=1$, for example, corresponds to an Einstein-de Sitter 
pure matter universe, or a de Sitter pure cosmological 
constant universe, or any model that instantaneously 
lies on the $w'=3w(1+w)$ line. Of course interpreting 
$q$ and $j$ as a Taylor expansion about the present 
expansion behavior would restrict their usage to $z\ll1$. 
Also note that while the scale factor can be viewed as 
a kinematical quantity (e.g.~no equation of motion need 
be specified, just the metric, to know how light is redshifted), 
this breaks down as soon as time dependence 
is explicit, e.g.~by parametrizing $q=q_0+q_1z$. Thus no 
advantage exists for such a representation over the dynamical 
phase space.

\section{Extending Dynamics \label{sec:extend}} 

We can now investigate whether the dynamics phase 
space $w$-$w'$ is useful for physical theories beyond canonical, 
minimally coupled scalar fields. This includes for 
modified gravity or other theories where the quantities $w$ 
and $w'$ are effective quantities, defined in terms of the deviation 
in the expansion rate from the matter dominated behavior, 
\beqa 
\delta H^2&\equiv&(H/H_0)^2-\om a^{-3} \\ 
w_{\rm eff}&\equiv&-1-\frac{1}{3}\frac{d\ln\delta H^2}{d\ln a}, 
\eeqa 
possibly distinct from any physical pressure or dark energy 
density. 

\begin{figure}[!htb]
\begin{center} 
\psfig{file=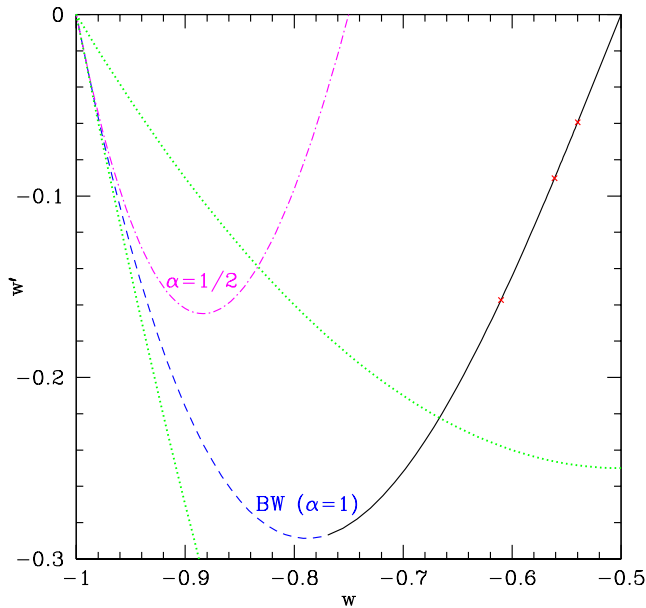,width=3.4in} 
\caption{Modifications to the Friedmann equation of the form 
$H^\alpha$ lie in the freezing regime, despite possibly not arising 
from a simple scalar field.  Moreover, they asymptotically approach 
$\lam$ along the lower boundary line $w'=3w(1+w)$.  The braneworld 
curve is shown solid to $z=0$, with x's indicating $z=1,2,3$. 
}
\label{fig:halpha} 
\end{center} 
\end{figure}

As already mentioned, phenomenological models such 
as $\delta H^2\sim H^\alpha$ \citep{dvaliturner} 
fit within the freezing picture and the 
specific freezing region of the phase space, as illustrated 
in Fig.~\ref{fig:halpha}. Note that the case $\alpha=1$ corresponds to 
the dynamics of an extra dimensional braneworld model \citep{dgp,ddg}; 
such models are discussed in more detail by \citet{koyama} in 
this volume.

Since the results of \S\ref{sec:dynwwp} were discussed in terms of 
canonical, minimally coupled fields, let us examine the 
extension to noncanonical or coupled dark energy. 

{\it k-essence:\/} If we remove the canonical nature of the 
scalar field Lagrangian that involves an additive term linear 
in the kinetic energy, we have a class of theories known 
as k-essence \citep{mukhanov,kessence}, with Lagrangians of the form 
\beq 
{\mathcal L}=V(\phi)\,F(X), 
\eeq 
where $X=(\partial_\mu\phi\,\partial^\mu\phi)/2$, i.e.~in the 
absence of spatial 
inhomogeneities $X$ is just the kinetic energy. Such models 
have some inspirations from field and string theory 
(for an overview see \citet[e.g.~][]{makler}), can describe phantom fields 
with $w<-1$, can have sound speeds less than the speed 
of light (hence affecting structure formation differently 
than quintessence) and can have attractor mechanisms 
to alleviate the fine tuning problem. 

Without further specifying the functions $V$ or $F$, it is 
difficult to say anything general about k-essence 
dynamics. Purely kinetic k-essence, where $V={\rm constant}$, 
does have phase space trajectories limited to one side or 
the other of the line $w'=3w(1+w)$ corresponding to 
constant pressure \citep{scherrerwwp,paths}.  
However kinetic k-essence 
can dynamically mimic (or be mimicked by) quintessence 
as long as the portion of the phase space trajectory of interest 
does not cross this line \citep{sen,deputter}. 

{\it Coupled dark energy:\/} The dark energy could in fact 
be not dark, that is it could interact non-gravitationally. 
From the dynamical perspective this creates an effective 
equation of state shifted from the bare one by the interaction 
term, e.g. 
\beq 
w_{\rm eff}=w-\frac{\Gamma}{3H}, 
\eeq 
where $\Gamma$ is the interaction appearing in the continuity equation 
\beq 
\dot\rho_w=-3H\rho_w (1+w)+\Gamma\rho_w, 
\eeq 
representing a decay/creation process for example. 
This was set forth in early work by \citet{turnerdk} and \citet{lin88s}. 
Such coupling will shift the trajectories in the $w$-$w'$ 
phase space, allowing for dynamics outside the thawing 
and freezing regions. Many different couplings, and their 
cosmological effects, have been considered; see, e.g., 
\citep{amendola,groexp,barnes}. 
However, concerns have been raised about the apparent 
strong effect of quantum corrections on fields coupled 
to matter \citep{doranjaeckel}. This can be avoided if one postulates that 
the potential considered is really an effective low energy 
potential that just happens to take on a simple form as 
a result of complicated quantum loop corrections to the 
(in turn necessarily complicated) classical potential; see 
the article by \citet{durrer} on low energy effective theories in this 
volume. 

{\it Scalar-tensor gravity:\/} Rather than coupling the dark 
energy to the matter sector of the Lagrangian, one could 
make the coupling to gravity nonminimal. These 
are scalar-tensor theories; see the article by \citet{francaviglia} 
in this 
volume. Coupling the quintessence field to the Ricci 
scalar, $R/(8\pi G)\to F(\phi)\,R$ in the action, these extended 
quintessence theories \citep{perrotta} can have varied dynamics depending 
on the form of $F$, along with an interesting attractor 
mechanism called the $R$-boost \citep{rboost}. For a model 
with a cosmological constant potential, requiring consistency 
with solar system tests drives the equation of state 
very close to $w=-1$ (within $10^{-4}$) and with dynamics 
representative of neither freezing nor thawing fields \citep{garbari}. 
For another approach, see \citet{nesseris}. 

{\it Model Zoo:\/} As the fertile imagination of children's author 
Dr.~Seuss envisioned an alphabet and animals ``On 
Beyond Zebra'', so has the intense interest in the dark 
energy mystery led to a zoo of models ``On Beyond $\lam$''. 
The merest glimpse of a small fraction of these includes: 
{\it oscillating\/} (see also {\it slinky\/}) models \citep{quigg,barenboim} 
with dynamics corresponding 
to a circle in phase space \citep{linosc}, {\it mocker\/} models 
that arc from matter like behavior to cosmological constant 
like behavior along curves of $w'=Cw(1+w)$ \citep{paths}, 
closely related to {\it quartessence\/} and {\it Chaplygin gas\/} models 
that attempt to unify dark matter and dark energy (see 
\citet{quart} for an overview), {\it skating\/} models that arc from free 
field behavior ($w=+1$), to cosmological constant like 
behavior along the curve $w'=-3(1-w^2)$, physically 
corresponding to a field moving across a constant potential 
\citep{sahlen,lincurv} (but also related to kinetic k-essence 
\citep{deputter}), and {\it wet fluid\/} \citep{holman} (equivalent 
to the sum of a constant $w$ component and a cosmological constant; 
cf.\ \S\ref{sec:complex}) or {\it leveling\/} \citep{paths} 
models that approach a cosmological constant as the density nears a 
limiting value and have parabolic tracks -- respectively 
$w'=3(1+w)(w-w_*)$ and $w'=-3(1+w)(w+w_*)$.

\section{Dynamics and Growth \label{sec:dyngro}} 

The dynamics of the accelerating component affects 
the growth of structure in the universe through the expansion 
rate. This provides a Hubble friction term opposing 
gravitational instability (e.g.~reducing the exponential 
Jeans growth in a static background to the power 
law growth in an expanding background). It also affects 
the matter source term $\om(a)$, i.e.~the evolution of 
the homogeneous matter density, through the expansion, 
but to the extent that dark energy remains smooth on the relevant 
scales it does not directly source growth. 
Canonical scalar fields are very light, $m\lesssim H$, so they 
remain smooth on scales less than the Hubble scale \citep{caldwellm}. 
Therefore, within general relativity, the growth effects of 
dark energy follow directly from the expansion effects discussed 
in this article. A highly accurate fitting formula 
for the linear growth can be given in terms of $\om(a)$ and 
$w(z=1)$ through the gravitational growth index formalism 
\citep{groexp,lincahn}: 
\beqa 
g(a)&\equiv&(\delta\rho_m/\rho_m)/a=e^{\int_0^a (da/a)[\om(a)^\gamma-1]} \\ 
\gamma&=&0.55+0.05\,[1+w(z=1)], 
\eeqa 
is accurate to 0.2\% compared to the numerical solution 
of the exact second order differential equation. Structure 
formation in general requires treatment of fully nonlinear 
growth through N-body numerical computations. Early 
work with dynamical quintessence included that of 
\citet{klypin,linjen}, with many following investigations. 

When the physics of the cosmic acceleration has a gravitational 
origin, or a dark energy component is not minimally 
coupled, additional terms enter into the growth, including 
new source terms such as from anisotropic stress 
and non-unity sound speed, and varying gravitational 
coupling. This breaking of the degeneracy between expansion 
effects and growth effects offers a promising window 
for identifying the fundamental physics, but is beyond 
the scope of this article; see, e.g., the review by 
\citet{linrop} for more details.

\section{Thawing Dark Energy \label{sec:thaw}}

Let us now return to the fundamental mode picture
of quintessential dynamics, presenting some new results 
on the specifics of determining the class of dark energy 
responsible for cosmic acceleration and the ability to zero 
in on characteristics within that class. 

While distinguishing the thawing class of dark energy from the freezing 
class would be a major accomplishment guiding us toward the fundamental 
physics behind dark energy, we can also examine thawing models in 
themselves.  These are among the best motivated physics, including 
radiatively stable PNGB pseudoscalar or axion models and familiar 
quadratic, quartic, and other renormalizable potentials. 

\subsection{Thawing Physics} 

Thawing models are defined by their departure from a cosmological 
constant-like state in the past to a dynamical, $w\ne-1$, behavior 
today.  This property of being frozen into a cosmological constant 
over much of the history of the universe makes this class difficult 
to distinguish from a cosmological constant without highly accurate 
cosmological data.  Indeed, current observations are almost wholly 
degenerate with the entire thawing region as defined in \citet{caldlin}, 
and if an effective, constant $w$ (e.g.\ a weighted average over the 
data sensitivity) is determined to equal $-1$ within 5\% then we still 
have essentially no information on whether this is truly a cosmological 
constant $\lam$ or any model in the entire half of the physical model 
space that is categorized as thawing. 

This challenge in uncovering the underlying physics makes this class 
useful as a testbed for the science reach of next generation experiments 
and for the role of phenomenological parameterization.  We will 
particularly be interested in, of course, distinguishing thawing models 
from $\lam$ and seeing dynamics such that $w(z)\ne w_{\rm const}$, but 
we also would learn physics more directly by verifying that the field 
started in a frozen state at early times and furthermore discerning 
its trajectory in phase space or at least its dynamical slope parameter 
$w'/(1+w)$. 

Recall from \citet{paths} that 
\beqa 
\frac{w'}{1+w}&=&2X+3(1+w) \\ 
&=&3\frac{1-Y}{1+Y}+3w, 
\eeqa 
where $X=\ddot\phi/(H\dot\phi)$ and $Y=\ddot\phi/V_{,\phi}$.  Thus, 
constraining the dynamical slope parameter $w'/(1+w)$ directly leads to 
information about the field acceleration, friction, and potential tilt 
terms in the Klein-Gordon equation of motion (\ref{eq:kg}). 

\subsection{Thawing Models} 

We begin by examining three parameterizations of thawing fields, 
comparing their behavior and constraints.  First is the standard 
parameterization of $w(a)=w_0+w_a(1-a)$, reviewed in 
\S\ref{sec:dynpar}.  If we choose $w_0+w_a=-1$, then we see that at 
early times ($a\ll1$), this possessed $w=-1$.  Furthermore, $w'=-aw_a$ 
so at early times $w'=0$; thus this parameterization can describe a 
thawing model.  However, we have handcuffed this parameterization 
by doing this, reducing it to a single parameter, rather than a 
model with two degrees of freedom, putting it at a disadvantage.  
It is basically restricted to the trajectory $w'=1+w$.  
Nevertheless, we will see that it is able to describe reasonably 
most thawing models.  The alternative is to retain the two parameters 
of $w_0$, $w_a$ but at the price of not matching a cosmological constant 
at early times; since cosmological data weights the recent universe 
more heavily, this is not a bad approximation.  The energy density of 
$w_a$ models is 
\beq 
\rho_w(a)=\rho_w\,a^{-3(1+w_0+w_a)}e^{-3w_a(1-a)}. 
\eeq 

The second parameterization $w'=F(1+w)$ is motivated by PNGB models, 
and is an excellent approximation to their dynamics \citep{caldlin}, 
with $F$ inversely proportional to the symmetry breaking energy scale $f$. 
For more on PNGB models, see \citet{frieman95,kalopersorbo,frieman06}. 
These have fields starting frozen on their $1+\cos(\phi/f)$ potential 
and, after 
the Hubble drag diminishes, are released to roll.  Due to the change 
from convexity to concavity of the potential, they can have interesting 
dynamics depending on the initial conditions.  We assume they are not 
fine tuned in the sense of starting very near the top of their potential, 
nor have they already rolled through the minimum and ascended the 
potential.  Eventually the field will oscillate around the minimum 
(which looks quadratic, i.e.\ $V\sim\phi^n$ with $n=2$, so the 
effective equation of state $w=(n-2)/(n+2)=0$ \citep{turner83} as 
long as the oscillation period is short compared to the Hubble time), 
acting like dark matter before vanishing as the field 
comes to rest at zero potential.  However, during the accelerating 
period, $w'=F(1+w)$ accurately describes the dynamics; the equation 
of state has two parameters, the current equation of state $w_0$ 
and the dynamical slope $F$, with 
\beq 
1+w=(1+w_0)\,a^F. 
\eeq 
The energy density of these models is 
\beq 
\rho_w(a)=\rho_w\,e^{[3(1+w_0)/F](1-a^F)}. 
\eeq 

For the third parameterization, we craft a new model specifically 
following the physics of thawing, called the algebraic thawing model: 
\beq 
1+w=(1+w_0)\,a^p\,\left(\frac{1+b}{1+ba^{-3}}\right)^{1-p/3}, \label{eq:alg}
\eeq 
with parameters $w_0$, $p$ ($b$ is fixed).  Let us justify this form.  
As the field is released from 
the cosmological constant state, still in the matter dominated era 
$t\sim a^{3/2}$, the dynamics is given by $X=3/2$, or $w'=3(1+w)$, 
as illustrated in \citet{caldlin}.  This implies that at early 
times $1+w\sim a^3$.  So far this is identical to the PNGB model 
with $F=3$.  To add some curvature into the trajectory in the 
phase plane $w$-$w'$, let us multiply this by a factor that bends 
the dynamics away from this line as the scalar field energy density 
becomes more important, say $\omw(a)^q$.  In fact, to preserve the 
early time behavior, this factor must go to a constant at early times, 
so we use $[a^3\omw(a)]^q$, which is indeed constant at early 
times when $w\to-1$.  The only problem with this is that the expression 
for the equation 
of state has become non-analytic.  Even if we approximate $\omw(a)$ 
by some fixed function, say $\Omega_\Lambda(a)$, then the equation of 
state is intertwined with the present energy density parameter $\omw$, 
or $\om$, rather than being an independent quantity.  For the final 
form we therefore replace the intruding density ratio -- in this one 
place -- with a constant $b=0.3$.  The equation of state is quite 
insensitive to this specific value, varying by less than 1\% as 
$b$ varies by 50\%; of course the value of $b$ is irrelevant as $a\to1$ 
and for $a\ll1$. 

The dynamics of the algebraic thawing model is 
\beq 
w'=(1+w)\,\left[3-\frac{3-p}{1+ba^{-3}}\right], 
\eeq 
and the energy density is 
\beq 
\rho_w(a)=\rho_w\,\exp\left[\frac{3(1+w_0)}{\alpha p}\left\{1-(\alpha a^3+\beta)^{p/3}\right\}\right], 
\eeq 
where $\alpha=1/(1+b)$, $\beta=b/(1+b)$. 

In a clever analysis, \citet{crittenden} came up with a similar model 
by analyzing a slow roll-like field expansion, assuming a particular 
combination $V_{,\phi}/[V(1+X/3)]$ can be Taylor expanded about the 
present.  After some approximations they take 
$1+w\sim a^p\,\Omega_\Lambda(a)^{1-p/3}$.  However, 
this form still entangles $w$ and the present matter density, and in 
fact a more exact solution of the field expansion equations works 
worse!  The basic problem is that even for thawing fields there is 
no reasonable slow roll or field expansion approximation.  Even for 
their less extreme model with $w_0=-0.8$, the field still traverses 
$\Delta\phi\sim 0.4M_P$.  The algebraic form Eq.~(\ref{eq:alg}) 
in fact gives more accurate equations of state for the cases they 
illustrate. 

\subsection{Discriminating Thawing} 

We can now use the $w_a$, PNGB, and algebraic models to examine the 
constraints, and parameter dependence of the constraints, from future 
data on the dynamics of quintessence.  For each model we have two 
equation of state fit 
parameters: $(w_0,w_a)$, $(w_0,F)$, or $(w_0,s)$, where $s=w'_0/(1+w_0)$ 
is the dynamical slope at present (just as $F$ is the dynamical slope, 
constant for all times).  From the estimation of these parameters 
(marginalizing over the matter density, in a flat universe, and other 
parameters such as the supernovae absolute magnitude), and 
their covariances, we can find the constraints on $w$ and $w'$ at any 
redshift, giving confidence contours in the $w$-$w'$ phase plane. 

For future data we consider Type Ia supernovae distances from $z=0-1.7$, 
with systematics, of SNAP quality (see, e.g., \citet{snap}), plus the 
reduced distance to the CMB last scattering surface, of Planck quality 
(0.7\% fractional precision).  The fiducial cosmology has $w_0=-0.9$ and 
present dynamical slope 1.5, and likelihoods are approximated as Gaussians 
in a Fisher information analysis.  The CMB data in fact has little leverage 
on the equation of state, because for all the thawing models the high 
redshift equation of state goes to a cosmological constant.  We have 
checked that adding baryon acoustic oscillation angular distance 
measurements at 1\% precision or a matter density prior of 0.005 (roughly 
mimicking weak gravitational lensing constraints) does little to improve 
the constraints. 

\begin{figure}[!htb]
\begin{center} 
\psfig{file=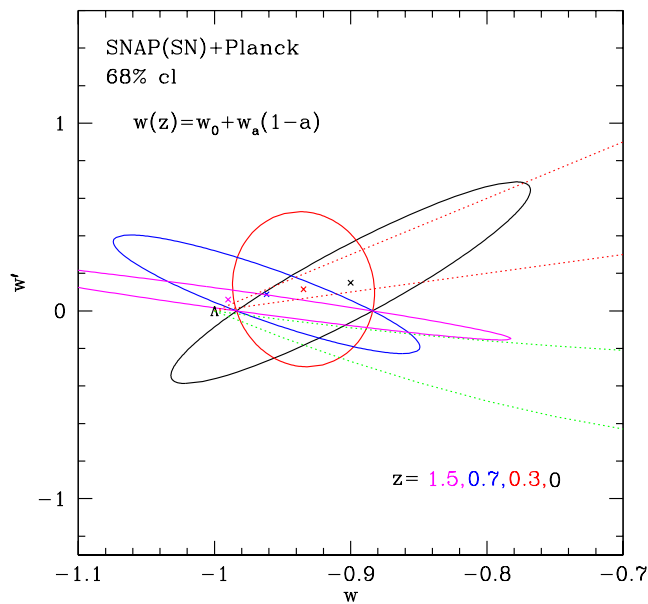,width=3.4in} 
\caption{Together with Figs.~\ref{fig:pngbwwpell}, \ref{fig:thawwwpell}, 
this figure for the $w_a$ thawing model illustrates constraints on the 
dynamical behavior of three thawing models at four redshift snapshots.  
While the $z=0$ behavior 
is poorly limited by future data, taking into account the dynamical 
history still allows distinction of the fiducial $w_0=-0.9$, $w'_0=0.15$ 
model from a cosmological constant and from the freezing class of physics. 
}
\label{fig:thawwawwpell} 
\end{center} 
\end{figure}

\begin{figure}[!htb]
\begin{center} 
\psfig{file=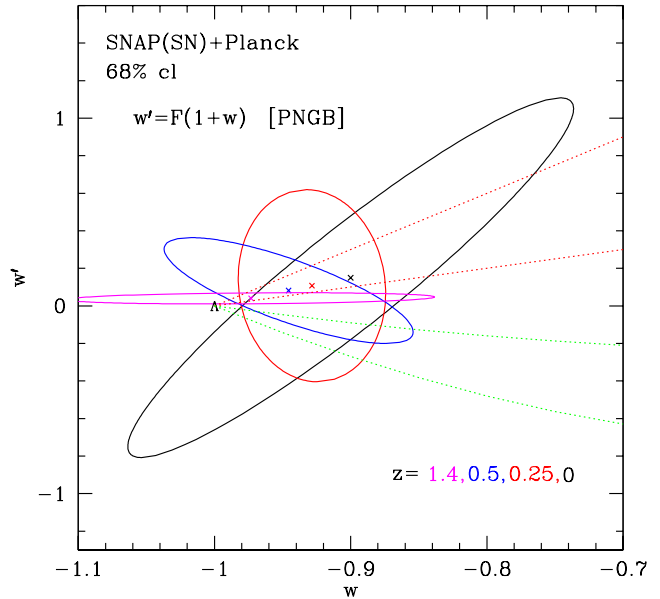,width=3.4in} 
\caption{As Fig.~\ref{fig:thawwawwpell}, for the PNGB thawing model. 
}
\label{fig:pngbwwpell} 
\end{center} 
\end{figure}

\begin{figure}[!htb]
\begin{center} 
\psfig{file=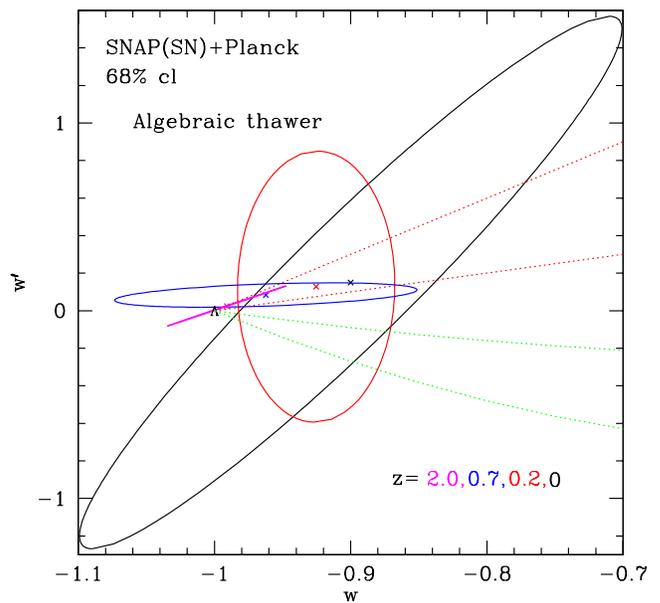,width=3.4in} 
\caption{As Fig.~\ref{fig:thawwawwpell}, for the algebraic thawing model. 
}
\label{fig:thawwwpell} 
\end{center} 
\end{figure}

Figures~\ref{fig:thawwawwpell}-\ref{fig:thawwwpell} show the $w$-$w'$ 
constraints for the three 
models at four redshifts.  We exhibit the 68\% confidence level 
contours at $z=0$, at the redshift where $w$ and $w'$ are decorrelated, 
giving vertical/horizontal ellipses, and at high redshift, $z\gtrsim1$. 
The phase space trajectory is marked by the x's at each of the four 
redshifts.  Note that the confidence contours vary between the models, 
especially when evaluated at the present, and this may lead to concerns 
about parameterization dependence.  However, as we will see, the 
qualitative answers to the important physical questions remain independent 
of the parameterization. 

While constraints on the present dynamical state, i.e.\ $w_0$ and $w'_0$, 
are relatively weak, in each of the parameterizations they are still 
sufficient to distinguish the fiducial model $w_0=-0.9$, $w'_0=0.15$ 
from a cosmological constant.  (Note this is despite the uncertainty 
$\sigma(w_0)\approx0.14$ from the algebraic thawer, the weakest model at 
$z=0$ -- one must take into account the contour orientation in the 
phase plane.) 
At $z=0$, however, the models cannot distinguish thawing from freezing, 
or from a constant equation of state $w_{\rm const}=-0.9$.  Using the 
information from throughout the dynamical history greatly improves 
the situation.  At some redshift, $z\approx0.2-0.3$ in the cases here, 
the dynamical variables $w$ and $w'$ decorrelate and the contours 
become vertical.  This gives the greatest distance between the constraint 
contour and the cosmological constant, showing clear distinction, and 
the intersection of the ellipse with the $w'=0$ axis also provides the 
minimal variance estimate on the instantaneous equation of state value. 
Such a decorrelation redshift is sometimes called a pivot redshift. 
Generically there can be more than one decorrelation redshift, and for 
the models where $w$ is not a linear function of the parameters we 
exhibit the contours 
at the second of these redshifts, $z\approx0.7-1.4$.  This provides a 
minimum variance estimate of the instantaneous time variation of the 
equation of state. 

Note that the confidence contours at each redshift are distinct from 
the cosmological constant, showing that future data can distinguish 
thawing models from $\Lambda$ (at least at $1\sigma$ for this fiducial 
cosmology). 
Furthermore, the early time contours (except in the $w_a$ case) 
distinguish the thawing model from models with constant equation of 
state, thus exhibiting the presence of dynamics.  The early time contours 
also draw away from the freezing region of the phase plane, so the data 
can indeed guide us to the correct class of physical origin.  These 
are all important physical insights that are not parameterization 
dependent.  Gains are more modest in zeroing in on a specific thawing 
model and these are more sensitive to parameterization.  At early times, 
the form of the algebraic thawer forces the contour to prefer a dynamical 
slope near 3.  However the PNGB and $w_a$ cases do not impose such 
preferences since the slope is a free fit parameter.  They do constrain 
$w'/(1+w)$ to a subset of the thawing region, rather than the full 
range of 1-3. 

It is heartening that the physical insights can be expected to be as 
clear as indicated, and not particularly dependent on the specific 
parameterization.  The issue of fitting the dynamical behavior of 
dark energy (especially when restricted to two parameters, as seems 
likely from realistic next generation data accuracy), is a fascinating 
one.  Use of a global parameterization like $(w_0,w_a)$ allows a 
good fit for models over the whole phase plane, but one can imagine 
that as we close in on the physical origin of dark energy, e.g.\ 
narrowing in on thawing models, we may move to more specific 
parameterizations such as the algebraic thawing model.  On the other 
hand, perhaps specific physical benchmark models, such as PNGB or 
motivated scalar field potentials, will then be of most use.

\section{Conclusion \label{sec:concl}} 

Dynamics, of quintessence and of the accelerating physics in general, 
can provide considerable insight into the nature of the new component 
or new physical law dominating our present universe.  Fundamental modes 
of the physics lead to well defined, distinct regions of $w$-$w'$ 
phase space that next generation cosmological probes will be able to 
test and distinguish.  Just as we build our physical intuition in 
early universe inflation with single field models leading to consistency 
relations, the fundamental modes of dark energy -- the quintessence 
of dynamics -- are a useful foundation. 

Model independent parametrization, with a strong physical basis, plays 
an important role, even if stringent constraints will be limited to 
two parameters such as the tilt from a cosmological constant, $1+w$, 
and a variation $w'$.  Nevertheless, this is as much as we expect from 
inflation as well, while for dark energy we have added complications due 
to the incomplete dominance of dark energy. 

Sensitivity to dynamics is a requirement to make progress in understanding 
the nature of cosmic acceleration.  Once we begin to zero in on a class 
of physics, model independence may give way to specific discriminating 
approaches such as the thawing analysis presented here.  Models for 
the equation of state which depend nonlinearly on the time variation 
parameter also possess minimum variance, or pivot, redshifts for the 
time variation, $z_{p'}$, and this may prove a useful tool. 

Dynamics alone, whether by its characterization or absence, will not 
fully solve the dark energy enigma.  The cosmic expansion history must 
be properly compared with the cosmic growth history to reveal extensions to 
gravitational physics or microphysics.  We have scarcely addressed this 
important subject 
here, nor have we said why in the presence of dynamics $\lam$ should not 
still exist, at a much larger energy density than the present, causing an 
abnegation of the universe we observe.  

Ten years passed from the time the 
basic physics and cosmology for the accelerating universe were in place 
until the first convincing observational evidence for its reality; since 
then another ten years of work on all fronts have passed. 
There is clearly still an enormous amount of exciting and challenging work 
ahead, and the answers, whatever they are and whenever they come, will 
revolutionize our understanding of gravitation, quantum physics, cosmology, 
and the fate of our universe.

\acknowledgments

This work has been supported in part by the Director, Office of Science,
Department of Energy under grant DE-AC02-05CH11231.

\end{document}